\documentclass[a4,aps,amsmath,floatfix,nofootinbib,10pt]{revtex4} 
\usepackage{graphicx} \usepackage{color} \usepackage{enumerate} 
\newcommand{\be}{\begin{equation}} \newcommand{\ee}{\end{equation}} 
\newcommand{\ba}{\begin{array}} \newcommand{\ea}{\end{array}} 
\newcommand{\bea}{\begin{eqnarray}} \newcommand{\eea}{\end{eqnarray}} 
\newcommand{\bdm}{\begin{displaymath}} 
\newcommand{\edm}{\end{displaymath}}

\begin{document}

\title{Dependence of persistence exponent on initial state}

\author{Prabodh Shukla}
\affiliation {Department of Physics, North Eastern 
Hill University, Shillong-793022, India} 

\begin{abstract} We examine persistence in one dimensional Ising model 
under zero temperature Glauber dynamics for random initial states with 
unequal fraction of up and down spins. We find the persistence exponent 
varies continuously with the fraction of up spins in the initial state. 
Apparently this feature has been overlooked in the studies so far. 
\end{abstract}

\maketitle

%%%\section{Introduction}

The one dimensional Ising model~\cite{ising} is an utterly simple model 
of cooperative behavior but it provides useful insight into several 
complex phenomena. A disordered state of Ising spins evolving under 
zero-temperature Glauber dynamics~\cite{glauber} coarsens~\cite{bray1} 
due to the competition between two stable states (all spins up or all 
down) at zero temperature. It is a slow stochastic process. Domain walls 
separating segments of up and down spins wander randomly and annihilate 
each other if they collide. Thus the density of domain walls decreases 
and consequently the average size of a domain increases with time $t$ 
(Monte Carlo steps) as $t^{1/2}$. Spins near domain walls show a wide 
range of flipping rates representing a distribution of relaxation times 
in the system. A finite fraction of spins in the system do not flip 
even once for a considerable length of time $t$. This surprising effect 
was noticed in mid nineteen nineties and is known as 
persistence~\cite{stauffer, krapivsky, derrida1}. Persistence is 
characterized by a persistence exponent $\theta$ such that the 
probability that a designated spin has not flipped up to time $t$ scales 
as $t^{-\theta}$ for large $t$. If a randomly chosen spin in the 
initial state has equal probability to be up or down then it was 
numerically observed that $\theta \approx 0.37$. This persistence 
exponent $\theta$ is apparently unrelated to any other exponent for the 
one dimensional Ising model. An analytic solution of the problem showed 
that $\theta=3/8=0.375$ exactly but the derivation of the exact 
result~\cite{derrida2} turned out to be surprisingly difficult. We may 
mention that the exact value $\theta=3/8$ is obtained under sequential 
dynamics. If parallel dynamics is used the persistent exponent 
$\theta_p$ is equal to $2 \theta=3/4$~\cite{menon}. The relationship 
$\theta_p=2 \theta$ follows from the bipartite nature of the one 
dimensional lattice. Therefore one may use either sequential or 
parallel dynamics for studying persistence and we choose parallel 
dynamics in the present study.

The new element in the present study is that we examine random initial 
states with an unequal fraction of up and down spins. In the following 
$c$ denotes the fraction of up spins in the initial state and $P(c;t)$ 
the corresponding persistence probability under parallel dynamics. As 
may be expected on account of the up-down symmetry of the Ising model, 
$P(1-c;t)$ and $P(c;t)$ are equal to each other, and so it suffices to 
focus on the case $c \le 0.50$. Persistence is a property of dynamical 
evolution of the system so it should not come as a surprise that the 
initial state plays a role in it. However, as far as we know, this point 
has been overlooked in the past despite an extensive study of the 
problem over an extended period. We do find that the persistence depends 
upon the quantity $c (1-c)$. Fig.1 shows the the persistence probability 
$P(c;t)$ vs. $t$ on a log-log plot for several representative values of 
$c$. A power-law behavior is clearly evident for each $c$. Persistence 
decreases fastest at $c=1/2$, and more slowly with decreasing $c(1-c)$. 
This is reasonable because in the limit $c \to 0$ or $c \to 1$ the 
initial state is invariant under the zero-temperature dynamics. In order 
to optimize the computer time, we studied a system of $10^6$ spins for 
$t \le 10^5$ Monte Carlo steps. Each MC step represents one update of 
the entire system under parallel dynamics. The data in Fig.1 is averaged 
over $10$ independent realizations of the initial state for each $c$. 
This seems adequate to infer the power-law behavior over several 
decades. Fig.2 shows the variation of the persistence exponent 
$\theta_p$ with $c$ over the range $0 \le c \le 1$ at intervals $0.05$. 
The continuous curve in Fig.2 is an aid to the eye. It is evidently 
difficult to obtain the $\theta_p(c)$ vs. $c$ curve analytically. We 
hope the numerical results presented here will motivate further work in 
this direction and better understanding of persistence and first-passage 
properties of non-equilibrium systems~\cite{bray2}.

\begin{figure}[ht] 
\includegraphics[width=0.75\textwidth,angle=0]{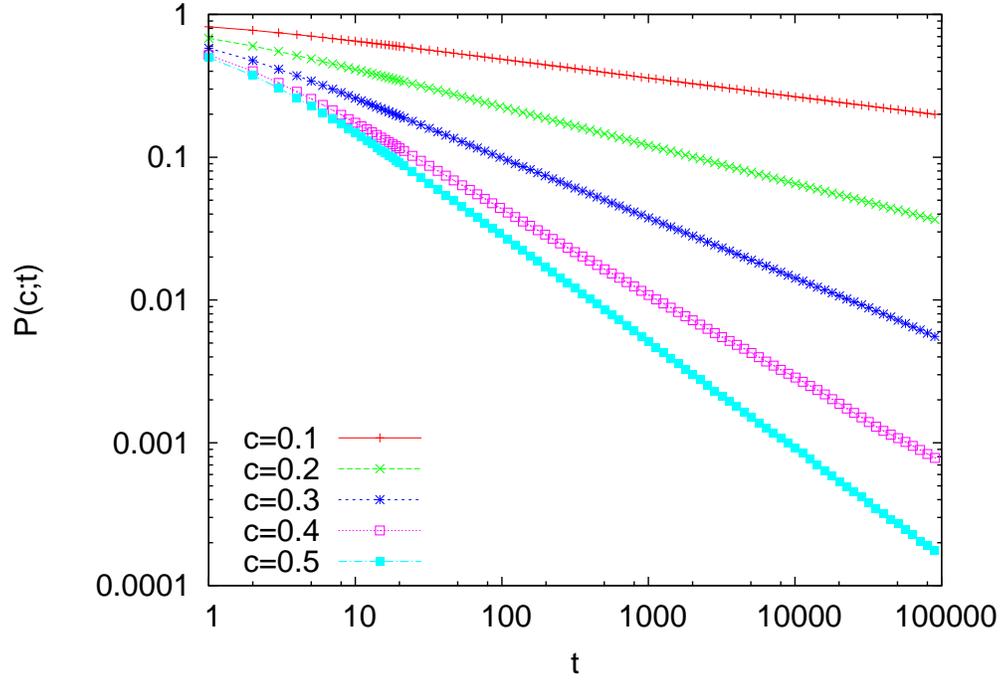} \caption{ 
Power-law decay of persistence probability $P(c;t)$ with time $t$ 
(number of MC steps). The power-law depends on the fraction $c$ of up 
spins in the initial state. } \label{fig1} \end{figure}

\begin{figure}[ht] 
\includegraphics[width=0.75\textwidth,angle=0]{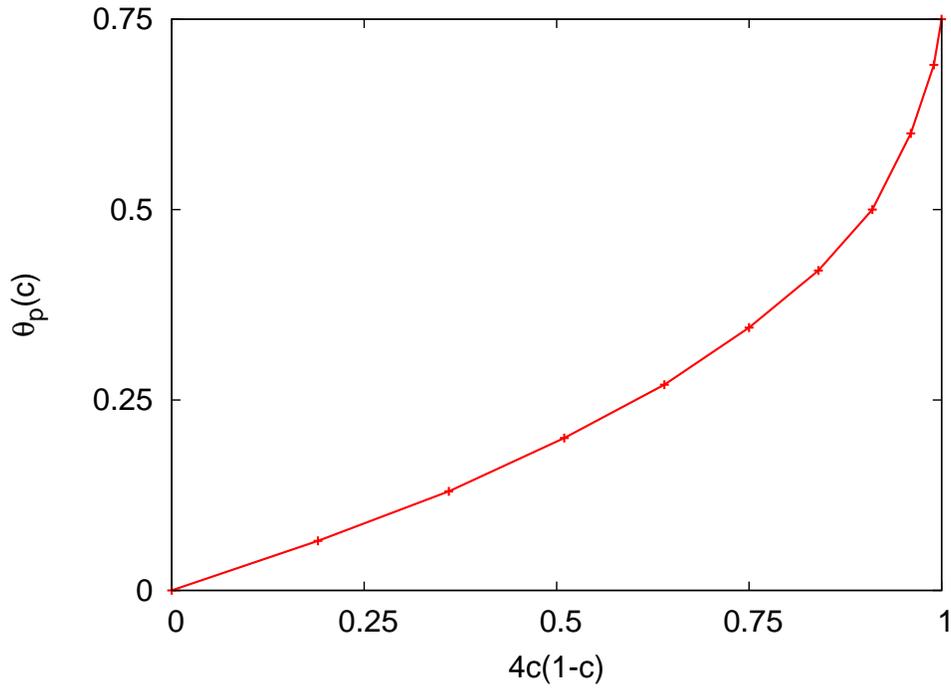} \caption{The 
dependence of the persistence exponent $\theta_p(c)$ on the fraction $c$ 
of up spins in the initial state.} \label{fig2} \end{figure}

\end{document}